\documentclass[aps,twocolumn,amsmath,amssymb,showpacs,prb,superscriptaddress]{revtex4-1}
\pdfoutput=1
\usepackage{epsf}
\usepackage{graphicx}

\begin{document}

\title{Evidence of momentum dependent hybridization in
Ce$_{2}$Co$_{0.8}$Si$_{3.2}$}

\author{P. Starowicz}
\email{pawel.starowicz@uj.edu.pl}
\affiliation{M. Smoluchowski
Institute of Physics, Jagiellonian University, Reymonta 4, 30-059
Krak\'{o}w, Poland}

\author{R. Kurleto}
\affiliation{M. Smoluchowski Institute of Physics, Jagiellonian
University, Reymonta 4, 30-059 Krak\'{o}w, Poland}

\author{J. Goraus}
\affiliation{Institute of Physics, University of Silesia,
Uniwersytecka 4, 40-007 Katowice, Poland}

\author{H. Schwab}
\affiliation{Universit\"{a}t W\"{u}rzburg, Experimentelle Physik
VII, Am Hubland, D-97074 W\"{u}rzburg, Germany}
\affiliation{Karlsruher Institut f\"{u}r Technologie KIT,
Gemeinschaftslabor f\"{u}r Nanoanalytik, D-76021 Karlsruhe,
Germany}

\author{M. Szlawska}
\affiliation{Institute of Low Temperature and Structure Research,
Polish Academy of Sciences, P.O. Box 1410, 50-950 Wroc\l{}aw,
Poland}

\author{F.~Forster}
\affiliation{Universit\"{a}t W\"{u}rzburg, Experimentelle Physik
VII, Am Hubland, D-97074 W\"{u}rzburg, Germany}
\affiliation{Karlsruher Institut f\"{u}r Technologie KIT,
Gemeinschaftslabor f\"{u}r Nanoanalytik, D-76021 Karlsruhe,
Germany}

\author{A. Szytu\l{}a}
\affiliation{M. Smoluchowski Institute of Physics, Jagiellonian
University, Reymonta 4, 30-059 Krak\'{o}w, Poland}

\author{I. Vobornik}
\affiliation{CNR-IOM, TASC Laboratory, SS 14, km 163.5, I-34149
Trieste, Italy}

\author{D. Kaczorowski}
\affiliation{Institute of Low Temperature and Structure Research,
Polish Academy of Sciences, P.O. Box 1410, 50-950 Wroc\l{}aw,
Poland}

\author{F. Reinert}
\affiliation{Universit\"{a}t W\"{u}rzburg, Experimentelle Physik
VII, Am Hubland, D-97074 W\"{u}rzburg, Germany}
\affiliation{Karlsruher Institut f\"{u}r Technologie KIT,
Gemeinschaftslabor f\"{u}r Nanoanalytik, D-76021 Karlsruhe,
Germany}

\begin{abstract}

We studied the electronic structure of the Kondo lattice system
Ce$_{2}$Co$_{0.8}$Si$_{3.2}$ by angle-resolved photoemission
spectroscopy (ARPES). The spectra obtained below the coherence
temperature consist of a Kondo resonance, its spin-orbit partner
and a number of dispersing bands. The quasiparticle weight related
to the Kondo peak depends strongly on Fermi vectors associated
with bulk bands. This indicates a highly anisotropic
hybridization between conduction band and 4f electrons - $V_{cf}$
in Ce$_{2}$Co$_{0.8}$Si$_{3.2}$.

\end{abstract}


\maketitle

\section{INTRODUCTION}

Cerium intermetallics are rich in exciting phenomena due to
various manifestations of the hybridization between a conduction
band and 4f electrons ($V_{cf}$) based on the Kondo
interaction~\cite{Stewart1984,Hewson1993}. Among the most peculiar
effects one may list the formation of heavy fermions,
unconventional superconductivity, quantum critical phenomena and
non-Fermi liquid behavior. Microscopic models for heavy fermion
(HF) systems assume very often that the strength of $V_{cf}$
hybridization is momentum independent~\cite{Weber2008}. On the
other hand, there are theoretical
considerations~\cite{Shim2007,Weber2008,Ghaemi2008} indicating a
strong momentum dependence of $V_{cf}$ with possible maxima and
nodes. This complex variation of the $V_{cf}$ amplitude is
expected to result from the symmetry of f-electrons, which
participate in the formation of the Kondo singlet. Therefore, the
issue of $V_{cf}$ anisotropy deserves more attention in the
experimental investigation of f-electron systems.

A typical spectroscopic manifestation of the Kondo effect is the
Abrikosov-Suhl resonance also called the Kondo resonance (KR)
which is a narrow high intensity peak in the spectral function
located close to the Fermi energy
($E_F$)~\cite{Malterre1996,Allen2005}. In case of Ce based
systems angle resolved photoemission spectroscopy (ARPES) was able
to determine a k-vector dependent intensity variation of the
KR~\cite{Andrews1995,Andrews1996,Garnier1997,Danzenbacher2005}. It
was reported that the increased KR intensity is found at normal
emission~\cite{Garnier1997} or at k-vectors related to bands
crossing $E_F$ (Fermi vectors)~\cite{Danzenbacher2005}. Recently,
it was shown for CeCoIn$_{5}$~\cite{Koitzsch2013} that the
f-electron peak intensity depends considerably on a band crossing
$E_F$, which was interpreted in terms of an anisotropic $V_{cf}$.
Optical spectroscopy~\cite{Burch2007} also reveals anisotropic
hybridisation for Ce-115 compound. Another effect of the
4f-conduction band hybridization is KR dispersion or more
precisely heavy quasiparticle dispersion, which has also been
found in Ce
systems~\cite{Andrews1996,Fujimori2006,Im2008,Koitzsch2008,Klein2011}.
Nevertheless, a complete image of momentum dependent hybridization
strength $V_{cf}$ with the occurrence of eventual nodes predicted
by theory~\cite{Shim2007,Weber2008,Ghaemi2008} remains a challenge
for an experiment so far.

The object of our investigations, Ce$_{2}$Co$_{0.8}$Si$_{3.2}$,
crystallizes with a derivative of the hexagonal AlB$_{2}$-type
structure with the $P6/mmm$ space group~\cite{Szlawska2011}. As
established from detailed magnetic susceptibility, electrical
resistivity and heat capacity measurements of single-crystalline
specimens, Ce$_{2}$Co$_{0.8}$Si$_{3.2}$ does not order
magnetically down to 0.4 K. Its physical behavior at low
temperatures is governed by strong \textit{f}-ligand
hybridization, leading to enhanced electronic contribution to the
specific heat [$C/T =$ 200 mJ$/$(mol$_{Ce}$K$^2$) at 0.4 K] and
Kondo-like temperature variations of the electrical resistivity
with the characteristic temperature $T_K$ of about 50
K~\cite{Szlawska2011}. The $\rho (T)$ dependencies are dominated
by broad maxima near 80 K, which manifest a crossover from
incoherent to coherent Kondo regime. Most interestingly, below 10
K, all the bulk characteristics of Ce$_{2}$Co$_{0.8}$Si$_{3.2}$
show non-Fermi-liquid (NFL) features that are compatible with
theoretical predictions for Griffiths phases. These should be
related to a disorder in Ce-Si sublattice found
recently~\cite{Szlawska2011}. So far, PES studies performed on
Ce$_{2}$CoSi$_{3}$ polycrystals delivered only angle-integrated
spectra showing a KR~\cite{Patil2010,Patil2012}.

In this paper, we present ARPES studies of the Kondo lattice
Ce$_{2}$Co$_{0.8}$Si$_{3.2}$ system. The spectra consist of
dispersions originating from surface and bulk states and a Kondo
peak related mainly to bulk states. A specific k-vector
dependence of KR indicates that certain bulk bands cross $E_F$
with much higher quasiparticle weight than others. These results
point to a momentum and/or band dependent anisotropic $V_{cf}$
hybridization in Ce$_{2}$Co$_{0.8}$Si$_{3.2}$.

\section{EXPERIMENT}
Single crystals of Ce$_{2}$Co$_{0.8}$Si$_{3.2}$ have been grown by
the Czochralski method and characterized as described
elsewhere~\cite{Szlawska2011}. The ARPES experiment was conducted
at the APE beamline of Elettra synchrotron~\cite{Panaccione2009}
with a SES2002 electron spectrometer. Prior to the photoemission
studies the crystals were oriented with a Laue method.
Subsequently, they were cleaved at a pressure of $2\cdot10^{-11}$
mbar exposing flat surfaces along the (10$\overline{1}$0)
crystallographic plane. The measurements were carried out with
linearly polarized radiation, typically at a temperature of 25 K.
The energy and wave vector resolution was fixed to 18 meV and 0.01
$\mathring{A}^{-1}$ respectively. The Fermi energy was determined
regularly on evaporated gold. Band structure calculations were
performed for stoichiometric Ce$_{2}$CoSi$_{3}$ using the scalar
relativistic version of the full-potential local-orbital (FPLO)
code~\cite{calc1} with the Perdew-Wang~\cite{calc2}
exchange-correlation potential. Additional correlations within the
LSDA+U approach were accomplished employing the Around-Mean-Field
scheme~\cite{calc3}. An irreducible wedge of the Brillouin zone
comprised 133 points.

\section{RESULTS AND DISCUSSION}
\subsection{Valence band studied with PES and FPLO}

The valence band of Ce$_{2}$Co$_{0.8}$Si$_{3.2}$ was investigated
by means of photoemission spectroscopy with the photon energies
(h$\nu$) corresponding to lower (h$\nu$=25 eV) and higher
(h$\nu$=40 eV) photoionization cross section for Ce 4f
electrons~\cite{YehLindau1985}. The spectra obtained by
integrating ARPES data for h$\nu$=40 eV and h$\nu$=25 eV are shown
in Fig. 1. Usually, one calculates the difference of the spectra
corresponding to these energies to approximate a contribution from
4f electrons. However, in the case of
Ce$_{2}$Co$_{0.8}$Si$_{3.2}$ the photoionization cross section
for Co 3d electrons almost doubles for the excitation energy
increase from 25 to 40 eV according to the estimates for free
atoms~\cite{YehLindau1985}. Hence, the contribution from both
orbitals, Ce 4f and Co 3d, is estimated by such a subtraction.

The peaks of the well screened f$_{5/2}^{1}$ and f$_{7/2}^{1}$
final states (Fig. 1) indicate a high hybridization between 4f and
conduction band electrons. The characteristic tail of the Kondo
peak f$_{5/2}^{1}$ is located near $E_F$ and its spin-orbit
partner f$_{7/2}^{1}$ at 275 meV. The broad peak at $\sim$ 2 eV
cannot solely be attributed to the weakly screened f$^{0}$ final
state. This is because the ARPES data (Fig. 2) reveal high
intensity dispersing peaks near this energy, which cannot
originate from highly localized and weakly hybridized f-electrons.
Thus, a large part of the spectral weight at $\sim$ 2 eV should be
attributed to Co 3d electrons. Density of states (DOS) obtained
theoretically (not shown) by means of the FPLO method with LSDA
approach indicate that Si 3p, Co 3d and Ce 4f dominate the
valence band. The contribution from Si 3p to the photoemission
spectra is less significant due to a very low photoionization
cross section at h$\nu$=25 and 40 eV.

\begin{figure}[h]
\begin{center}
\includegraphics[width=2.6in]{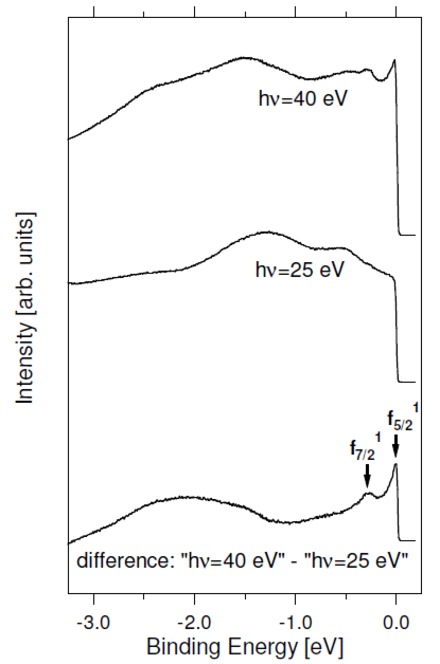}
\caption{ Photoemission Spectra obtained with photon energies of
h$\nu$=40 eV and 25 eV resulting from angle integration of the
ARPES data collected along the $\overline{\Gamma}$ -
$\overline{Y}$ direction of Ce$_{2}$Co$_{0.8}$Si$_{3.2}$ and their
difference.}
\end{center}
\end{figure}

Further insight into the electronic structure of
Ce$_{2}$Co$_{0.8}$Si$_{3.2}$ is given by band mapping with ARPES.
The path scanned in the reciprocal space is shown as a green
dashed curve inside the three-dimensional (3D) Brillouin zone (BZ)
[Fig. 2 (a)]. This curve is part of a large circle in the
$\Gamma$-A-L-M plane, for which the perpendicular to the surface
component of the wave vector (k$_{\perp}$) is unknown. For surface
states the problem is reduced to two dimensions and the band
structure is scanned along $\overline{\Gamma}$-$\overline{Y}$,
which is drawn with a green dashed straight line in the surface
Brillouin zone (SBZ). As the considered ARPES data do not deliver
any information about the real value of k$_{\perp}$ both 2D and 3D
bands will be described in the SBZ for simplicity.

\begin{figure*}[!]
\includegraphics[width=\textwidth]{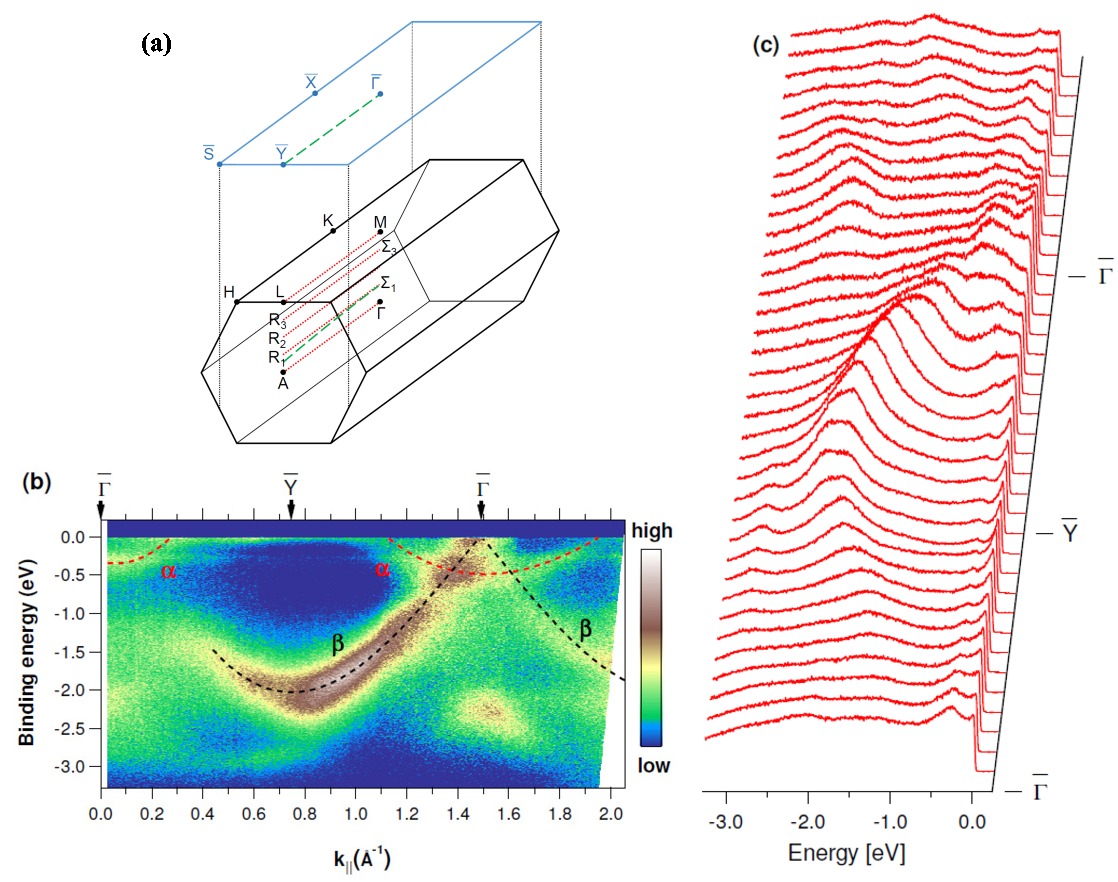}
\caption{(a) Three-dimensional Brillouin zone (black) and surface
Brillouin zone (blue) along (10$\overline{1}$0) plane for
Ce$_{2}$Co$_{0.8}$Si$_{3.2}$ with marked high symmetry points.
Dashed (green) line indicates a possible path in the reciprocal
space along which ARPES spectra were collected. Dotted (red)
lines represent paths $\Gamma$-A, $\Sigma_{1}$ - $R_{1}$,
$\Sigma_{2}$ - $R_{2}$, $\Sigma_{3}$ - $R_{3}$ and M - L used for
band structure calculations. (b) ARPES spectra for
Ce$_{2}$Co$_{0.8}$Si$_{3.2}$ obtained with the photon energy
h$\nu$=40 eV along $\overline{\Gamma}$ - $\overline{Y}$ direction
shown as intensity map and (c) energy distribution curves. Dashed
lines are guides to the eye indicating $\alpha$ and $\beta$
bands. The temperature of the sample was 25 K.}
\end{figure*}

The measurements were performed with h$\nu$=40 eV at a temperature
of 25 K, which is below the coherence temperature estimated to be
T$_{coh}\sim$ 80 K. KR and spin-orbit partner are not dispersing
but vary in intensity. The other peaks reveal dispersions. In
particular, $\alpha$ and $\beta$ bands (Fig. 2) clearly cross
$E_F$. The electron pocket ($\alpha$) observed around the
$\overline{\Gamma}$ points is more shallow near k$_{\parallel}$=0
than at k$_{\parallel}\sim$ 1.5 $\mathring{A}^{-1}$. The locations
in the reciprocal space scanned by ARPES for k$_{\parallel}$=0 and
k$_{\parallel}$=1.5$\mathring{A}^{-1}$ do not share the same out
of plane component of the wave vector - k$_{\perp}$ according to
the free electron final state model~\cite{Huefner}. Hence, 3D
states exhibit different dispersions for the equivalent
k$_{\parallel}$ but different k$_{\perp}$ in neighboring BZs.
Therefore, the $\alpha$ pocket is considered to be a 3D band of
bulk origin.

The $\beta$ band crossing $E_F$ at $\overline{\Gamma}$ is well
visible at k$_{\parallel}\sim$ 1.5 $\mathring{A}^{-1}$ in contrast
to k$_{\parallel}\sim$ 0 $\mathring{A}^{-1}$. This parabolic band
exhibits the same dispersion in both BZs and is therefore regarded
as a surface state. It has the highest spectral intensity of all
features in the spectrum, which is expected for surface states at
the photon energy of h$\nu$=40 eV. This is due to the very small
mean free path of electrons for this energy~\cite{Zangwill}. The
difference in the intensity at k$_{\parallel}\sim$ 1.5
$\mathring{A}^{-1}$ and k$_{\parallel}\sim$ 0 $\mathring{A}^{-1}$
may be attributed to matrix element effects, which favor certain
bands with a higher photoemission cross section, depending on the
geometry of the experiment, k-vector and probed BZ. This may
result in suppression or disappearance of particular bands. In
fact, the $\beta$ band looses intensity when approaching
k$_{\parallel}$=0 $\mathring{A}^{-1}$. At this wave vector the
experiment is realized in a pure $\pi$-polarization, which
excludes bands with odd parity with respect to the experimental
mirror plane. Due to this we cannot conclude whether this band
crosses $E_F$ near k$_{\parallel}$=0 $\mathring{A}^{-1}$ or bends
back to higher energies.

\begin{figure}[!]
\includegraphics[width=\columnwidth]{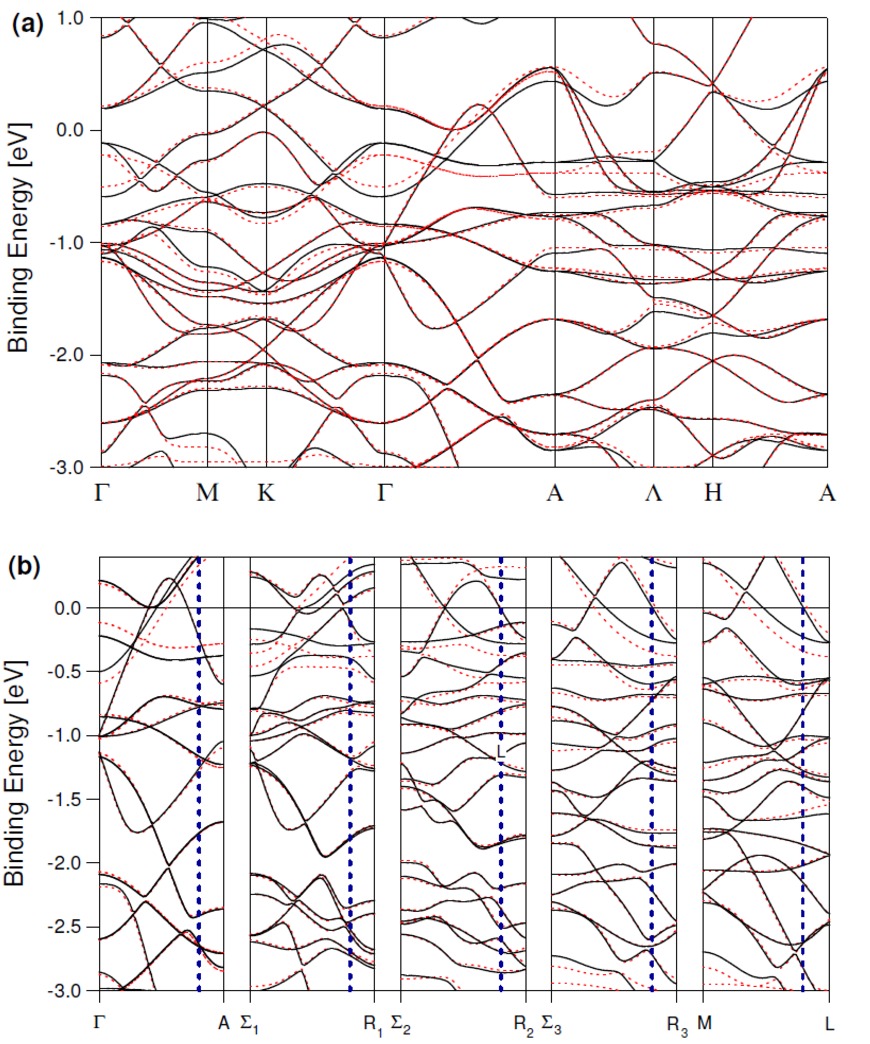}
\caption{ Theoretical dispersions obtained for bulk
Ce$_{2}$CoSi$_{3}$ by means of FPLO with LSDA+U approximation for
U$_{4f}$ =6 eV along (a) a standard path connecting high symmetry
points and (b) for the directions parallel to $\Gamma$-A. The high
symmetry points are defined in Fig 2 (a). The vertical dashed line
denotes k$_{\parallel}$=0.6 $\mathring{A}^{-1}$ with the highest
KR. Solid (black) and dashed (red) lines show dispersions
corresponding to spin up and down directions respectively.}
\end{figure}

Band structure calculations using the FPLO method (Fig. 3) are
helpful for a further interpretation of the ARPES results.
Theoretical dispersions for bulk Ce$_{2}$CoSi$_{3}$ are drawn
along a conventional path in the reciprocal space [Fig 3 (a)] and
for a set of directions parallel to $\Gamma$-A [Fig 3 (b)]. The
real path for the ARPES studies should be located close to one of
the latter directions [see also Fig. 2 (a)]. To consider different
strength of correlation the calculations were performed with the
Ce 4f correlation parameters $U_{4f}$ = 0 (not shown) and $U_{4f}$
= 6 eV. The calculations yield more bands, than observed in the
experiment. A certain number of them cannot be seen because of
unfavorable matrix elements. The electron pocket $\alpha$ found
experimentally near $\Gamma$ has its counterpart in the
calculations for $U_{4f}$ = 6 eV along $\Gamma$-A and
$\Sigma_{1}$-R$_{1}$. It reaches higher binding energy along
$\Gamma$-A and is a 3D state. On the other hand, the exact $\beta$
dispersion is not found in the calculations. This confirms the
surface state origin of this band. The other bands recorded with a
weaker intensity (Fig. 2) have corresponding dispersions in the
calculations with both $U_{4f}$ = 0 and 6 eV. The theoretical band
structure consists of two components related to opposite spin
direction. This is due to a magnetic ground state predicted by
calculations, which is however not found in the experiment. This
discrepancy can be explained by the fact that in the calculations
the total energy difference between magnetic and non-magnetic
ground state is so low that even at low temperatures the thermal
excitations would destroy long range magnetic order. Moreover,
Kondo scattering present in Ce$_{2}$Co$_{0.8}$Si$_{3.2}$ should
screen magnetic interactions.

\subsection{k$_{||}$-dependence of Kondo resonance intensity}

ARPES data provide a direct evidence of KR intensity variation
with k$_{\parallel}$, while its dispersion is not found at T=25 K,
a temperature well below the coherence temperature of the Kondo
lattice, T$_{coh} \sim$ 80K. Energy distribution curves (EDCs) for
exemplary k-vectors from Fig. 2 (b),(c) are shown in Fig. 4. The
height of the KR (blue arrow) was estimated with respect to the
background located between KR and the spin-orbit splitting peak
after proper normalization of the EDCs. The first surprising fact
is that the Kondo peak intensity is not directly correlated with
the Fermi wave vectors found in the same experiment. $E_F$
crossings at k$_{\parallel}$=0.3, 1.1 and
1.9~$\mathring{A}^{-1}$, which are related to the most prominent
bulk state, namely the $\alpha$ electron pocket, exhibit a
relatively low intensity Kondo peak. On the other hand, the
highest intensity KR is found around k$_{\parallel}$=0.6
$\mathring{A}^{-1}$, where bands crossing the Fermi energy are
not directly observed. The ARPES measurements were performed a
few times to exclude the effects of surface quality and KR
intensity dependence on k is reproducible.

\begin{figure*}[]
\includegraphics[width=5.2in]{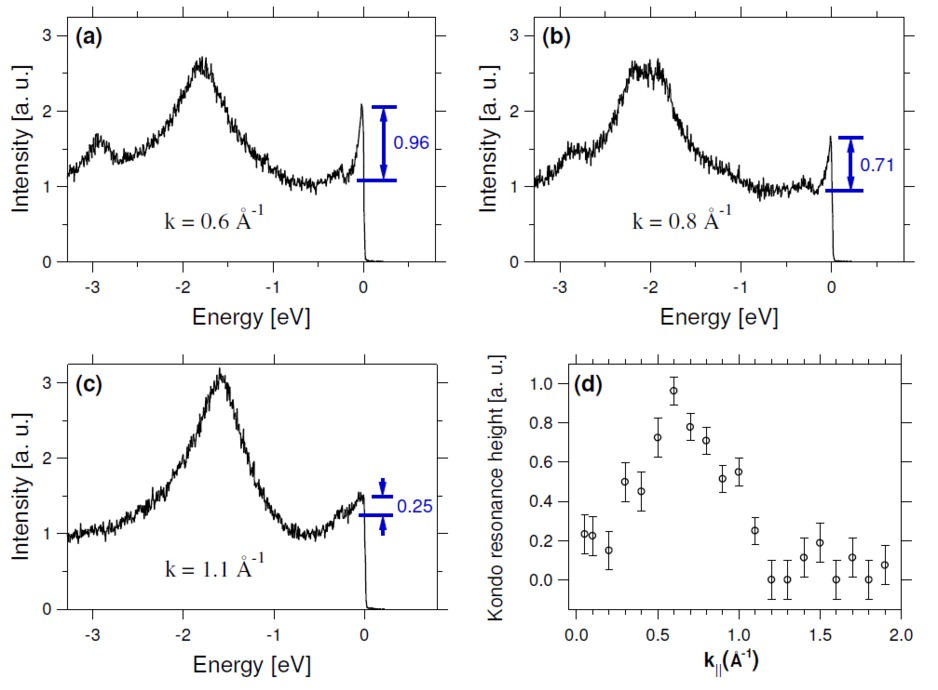}
\caption{ Energy distribution curves extracted from the ARPES
data (h$\nu$=40 eV) in Fig. 2 shown for (a) k$_{\parallel}$=0.6
$\mathring{A}^{-1}$, (b) k$_{\parallel}$=0.8 $\mathring{A}^{-1}$
and (c) k$_{\parallel}$=1.1 $\mathring{A}^{-1}$ with determined
KR height. Panel (d) presents KR height dependence on
k$_{\parallel}$.}
\end{figure*}

It is known that the KR may have a much higher intensity than its
parent band, which crosses $E_F$~\cite{Danzenbacher2005}.
Therefore, it is possible that the band responsible for the high
KR intensity is not visible. Bulk 4f electrons are known for a
stronger hybridization with the valence band electrons as compared
to surface states~\cite{Iwamoto1995} and they may form a high
intensity KR. Hence, the large KR should be assigned to a band,
which is not seen in the experiment and should be a bulk state.
This experimental material points to the main conclusion of this
letter, namely a large variation of the Kondo peak intensity along
Fermi surface (FS). KR is much higher for a weak intensity band at
k$_{\parallel}$=0.6 $\mathring{A}^{-1}$ than for the well visible
$\alpha$ pocket. Thus, taking into account band intensities one
may conclude that the real KR height variation is even much
larger. It is also noteworthy that the intensity of the KR is
lower for k$_{\parallel}$=0.9~$\mathring{A}^{-1}$ than for
k$_{\parallel}$=0.6~$\mathring{A}^{-1}$, wave vectors that are
equivalent in the SBZ. Due to the circular path of the ARPES
experiment mentioned before these k$_{\parallel}$ vectors
correspond to different values of k$_{\perp}$ and thus different
locations in the BZ. Inequality of KR height in these places is
consistent with the 3D nature of the related states.

In order to assign the highest intensity KR to a particular band,
the results of band structure calculations are shown along
$\Gamma$-A, M-L and the paths in between for LSDA+U approximation
with U$_{4f}$=6 eV [Fig. 3 (b)]. The wave vector 0.6
$\mathring{A}^{-1}$ corresponding to the highest KR is highlighted
with a dashed line. Actually, an $E_F$ crossing at 0.6
$\mathring{A}^{-1}$ appears for almost all considered paths
parallel to $\Gamma$-A. It should be stressed that in the case of
each considered path there are more E$_{F}$ crossings but the
highest intensity KR is observed only at the unique wave vector
in ARPES. The KR absence may also be explained by unfavorable
photoionization cross section for certain bands. Nevertheless,
the main thesis is based on the experimental spectra, namely on
the comparison of KR height for the $\alpha$ band and at
k$_{\parallel}$=0.6~$\mathring{A}^{-1}$.

Ce$_{2}$Co$_{0.8}$Si$_{3.2}$ was investigated in the coherent
state. Although, the f-state dispersion was not found directly in
the experiment, we assume that f-electron related quasiparticles
are present. The high intensity KR peak should refer to a high
quasiparticle spectral weight, Z(k) for a given momentum. Its
considerable variation along FS indicates also the momentum
dependence of $V_{cf}$. Such a situation was proposed in the
theoretical considerations for a Ce based cubic
system~\cite{Ghaemi2008}. These predicted strongly anisotropic
$V_{cf}$ resulting in a variation of Z(k) along the Brillouin
zone. In the cited report a high quasiparticle weight (Z(k)
$\sim$ 1) was found only in very small areas of FS along [001]
and equivalent symmetry directions. According to the predictions,
these regions can be easily missed by ARPES due to their very
limited size in k-space. Although similar calculations are not
realized for Ce$_{2}$Co$_{0.8}$Si$_{3.2}$ our ARPES results
support the theoretical predictions qualitatively giving the
first evidence of strongly anisotropic $V_{cf}$ in
Ce$_{2}$Co$_{0.8}$Si$_{3.2}$ system.

\section{CONCLUSION}
In conclusion, we investigated the band structure of
Ce$_{2}$Co$_{0.8}$Si$_{3.2}$ by means of ARPES and FPLO
calculations with LSDA+U approximation. Out of a larger number of
bands theoretically predicted for bulk Ce$_{2}$CoSi$_{3}$ only
some can be seen in the experiment. ARPES data reveal a bulk
electron pocket near the $\Gamma$ point ($\alpha$), which is found
in the calculations with U$_{4f}$=6 eV. The band $\beta$
exhibiting the highest intensity is interpreted as a surface
state. The main contributions from Ce 4f electrons to the
photoemission spectra are the Kondo resonance (KR) at $E_F$
associated with the f$_{5/2}^{1}$ final state and the peak related
to f$_{7/2}^{1}$ final state observed at 275 meV. Both peaks are
non-dispersing but their intensity varies as a function of the
wave vector. In particular, large maximum of KR is attributed to a
specific Fermi vector of a bulk band, whereas the other observed
bands from the bulk or the surface are characterized with a
medium or low quasiparticle weight at Fermi vectors. This
represents the ARPES evidence of a momentum dependent
hybridization between Ce 4f and conduction band electrons in
Ce$_{2}$Co$_{0.8}$Si$_{3.2}$.

\section*{ACKNOWLEDGMENTS}

This work has been supported by the Ministry of Science and Higher
Education in Poland within the Grant no. N N202 201 039. A part of
the measurements was carried out with the equipment purchased
thanks to the European Regional Development Fund in the framework
of the Polish Innovation Economy Operational Program (contract no.
POIG.02.01.00-12-023/08). HS, FF and FR acknowledge the support by
the DFG through FOR1162. J.G. acknowledges the financial support
from the National Science Centre (NCN), on the basis of Decision
No. DEC-2012/07/B/ST3/03027. We acknowledge technical support by
F. Salvador (CNR-IOM).

\end{document}